\documentclass[12pt]{article}
\usepackage{aaspp4}
\begin{document}

\title{    Confusion errors in astrometry and counterpart association}
\author{   David W. Hogg\altaffilmark{1}}
\affil{    Institute for Advanced Study, Olden Lane, Princeton NJ 08540\\
           hogg@ias.edu}
\altaffiltext{1}{Hubble Fellow}

\begin{abstract}
In very crowded fields, the modulation of the background by the sea of
unresolved faint sources induces centroid shifts.  The errors increase
with the number of sources per beam.  Even the most optimistic
simulations of imaging data show that position errors can become
severe (on the order of the beam size) at flux levels at which images
contain 1/50 to 1/15 sources per beam, depending on the slope of the
number-flux relation $d\log N/d\log S$.  These problems are expected
to be significant for recent observations of faint submillimeter
sources and may be the reason that some sources appear to lack optical
counterparts.
\end{abstract}


\section{Introduction}

Fainter is (usually) better when it comes to star and galaxy counts.
However, there are fundamental limits to faint imaging from confusion
which cannot be overcome by increasing exposure times alone.  The sea
of unresolved sources fainter than the detection limit creates a noise
in the sky, which does not improve with more data.

Many of the issues associated with this confusion noise have been
discussed before (eg, Scheuer 1957; Condon 1974; Franceschini 1982;
Hacking \& Houck 1987; Barcons 1992), however the large number of
present-day observations that are or soon will be pushing this
confusion limit suggests a new discussion.  In particular, in recent
years there has been a concerted effort to produce very deep,
multi-wavelength studies of blank sky in order to identify
extragalactic sources as comprehensively as possible.  These studies
have been very successful, identifying populations of radio-,
submillimeter-, infrared-, visual- and x-ray-bright galaxies and
associating them with their counterparts in other bands (Djorgovski et
al 1995; Williams et al 1996; Hogg et al 1996; Rowan-Robinson et al
1997; Richards et al 1998; Hughes et al 1998; Barger et al 1998; Eales
et al 1999; Aussel et al 1999; Elbaz et al 1999; Gardner et al 2000;
Brandt et al in preparation; Dickinson et al in preparation).  Some of
the faintest sources in some of the most crowded fields (in the sense
of number of sources per resolution element) have not shown clearly
distinguished counterparts at other wavelengths (Hughes et al 1998;
Smail et al 1998; Barger et al 1999b).  This raises the question
``could confusion be playing a role?''

This paper is a first attempt at characterizing position shifts due to
confusion in astronomical images.  Simulated images of crowded fields
are presented, made in the most optimistic way: no photon noise, a
perfectly understood gaussian point spread function or beam shape,
pointlike sources, a power-law number--flux relation of known slope,
and no angular clustering.  Even with these optimistic inputs, the
resulting images show that it is impossible to accurately measure
positions and fluxes of sources that are more than an order of
magnitude brighter than the flux level corresponding to one source per
beam (a beam being one resolution element in the image).  Recent work
making use of the limit ``one source per beam'' (eg, Blain et al 1998)
is therefore overly optimistic.

The standard rule-of-thumb is that confusion becomes important at 1/30
of a source per beam.  It is possible to get information from the
statistics of the background noise fainter than the level of 1/30
source per beam (eg, Scheuer 1957; Condon 1974), but in terms of
identifying and measuring individual sources, 1/30 is regarded as the
limit.  This paper tests the rule-of-thumb for confusion-induced
astrometry errors, which are particularly important for deep,
multi-wavelength studies, in which counterpart identification across
multiple data sets is important.  Astrometric shifts due to confusion
have been predicted and observed in the context of microlensing data
(Goldberg 1998; Goldberg \& Wo\'zniak 1998) and are expected to limit
future stellar astrometry experiments (Yu et al 1993; Rajagopal \&
Allen 1999).

For the purposes of this paper a ``beam'' is taken to be the solid
angle of the $1\,\sigma$-radius circle of the gaussian point spread
function, or $\Omega_{\rm beam}=\pi\,\sigma^2$.  Note that for a
gaussian, $\sigma\approx \theta_{\rm FWHM}/2.35$ where $\theta_{\rm
FWHM}$ is the angular full width at half-maximum of the point spread
function.  The number of sources per beam $s/b$ at a given flux level
$S$ is the integrated number of sources $N(>S)$ in an image brighter
than flux $S$ divided by the number of beams in the solid angle of the
image or $(\Omega_{\rm image}/\Omega_{\rm beam})$.

\section{Method and analysis}

\begin{figure}
\plotone{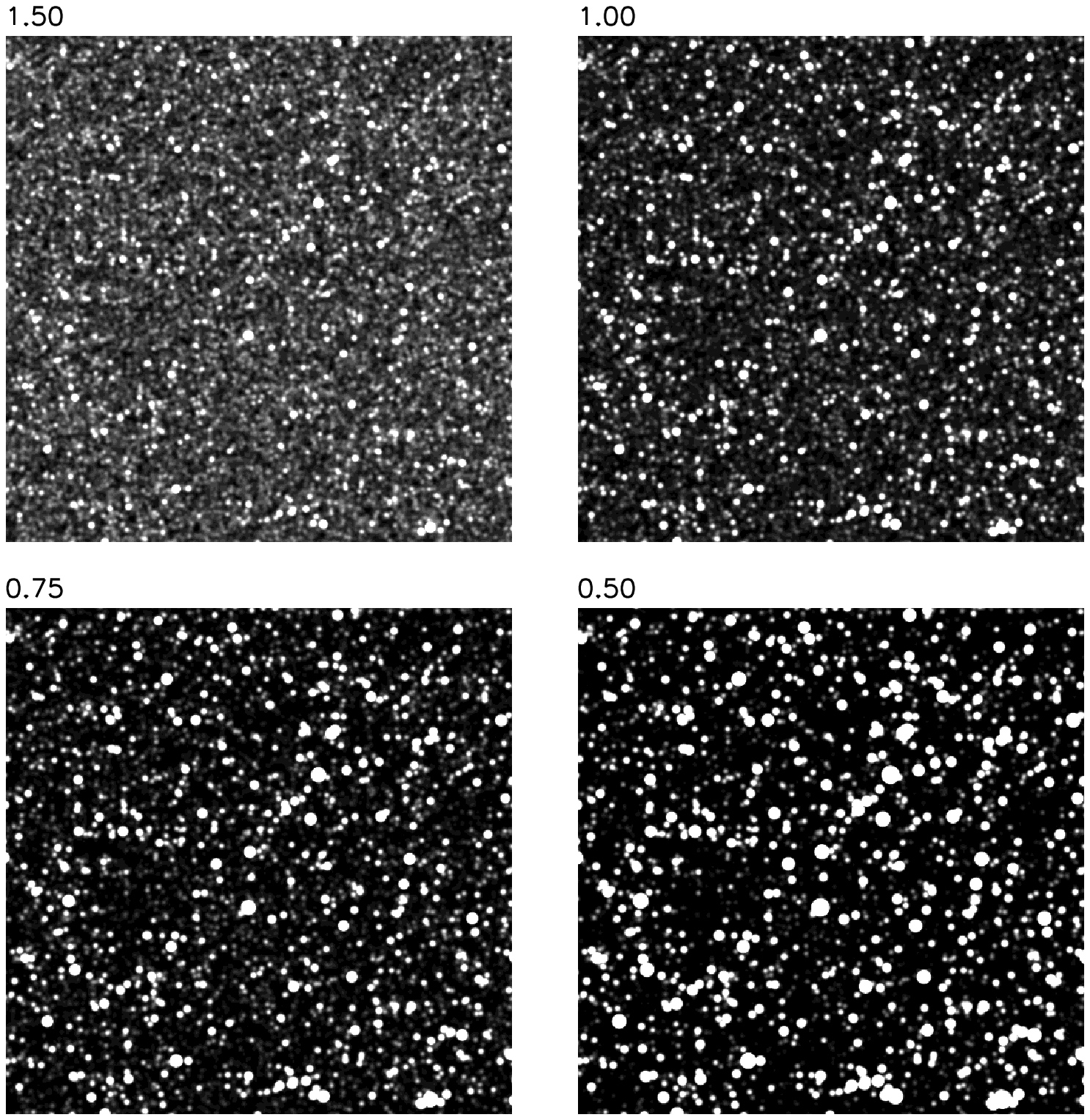}
\caption[ ]{ The four artificial images, labeled by the number-count
exponent $\beta$ (see text for definition).  The images are stretched
so that a source at the $s/b=1/30$ source per beam level appears the
same in all images.}
\label{fig:images}
\end{figure}

Four $512\times 512$~pixel artificial images were made with sources
randomly distributed over the image (and in fact over an area larger
than the image so that image edges are realistic).  The sources were
randomly assigned fluxes in power-law distributions $d\log N/d\log
S=-\beta$ where $\beta=1.50$, $1.00$, $0.75$, and $0.50$, one value of
$\beta$ per image.  Positions were randomly assigned and were not
quantized onto the pixel grid.  The point-spread function was chosen
to be perfectly gaussian with $\theta_{\rm FWHM}=4$~pixels, so that it
is well sampled.  Each image contains $\Omega_{\rm image}/\Omega_{\rm
beam}=2.88\times 10^4$~beams.  The artificial source catalogs were
truncated at $s/b=3$~sources per beam, ie, at much higher angular
density than the $s/b\sim 1/30$~sources per beam rule-of-thumb.  The
four artificial images are shown in Figure~\ref{fig:images}

No noise was added to the images; the sources are not extended beyond
the gaussian beam shape; and the sources were not given any angular
clustering.  The artificial images represent high optimism.

The background levels in the four images were fit by sigma-clipping
(ie, iteratively removing outlier pixels) at 3-sigma and they were
subtracted from the images.  Although the input background levels were
zero, the fit levels are above zero for all of the images because of
the integrated flux from all the unresolved sources.  In almost all
real observations but especially at wavelengths longer than the
near-ultraviolet (ground-based) or visual (space-based), images
contain large DC levels from sky emission or telescope thermal
emission so the true background (or, more accurately, foreground) is
unknown and must be fit by a procedure similar to the sigma-clipping
used here.  For $\beta\geq 1.0$, the background levels do not
converge, in the sense that the background light is dominated by the
faintest sources, and in these artificial images the background level
is just set by the depth to which the artificial source catalogs are
simulated.  However, experiments involving changing this depth show
that the confusion noise or level of background modulation in the
artificial images has converged.

Note that the sigma-clipping background estimation technique is
equivalent to (although less subjective than) estimating the
background from regions of the image that appear empty or blank.

At each location of a source in the artificial source catalog, a
centroid is found in the artificial image in a box of side length
$2\,\theta_{\rm FWHM}$ centered on the artificial source location in
the catalog.  These centroids are what will be referred to as the
``measured'' positions.

The catalog was trimmed to only ``isolated sources'': The source
positions measured by centroiding are in general shifted from the true
positions in the artificial source catalog; sometimes this is because
there is a brighter source nearby that is blending with the fainter
source of interest, sometimes this is just because the source is
projected onto the roiling sea of unresolved, fainter sources.  Only
the latter effect is properly an effect of confusion noise.  For this
reason, sources with measured positions closer than $2\,\theta_{\rm
FWHM}$ from a brighter source were dropped from the analysis that
follows.  This restriction to ``isolated'' sources (excluding faint
sources found to be near brighter sources) removes many of the largest
deviations of measured from true positions, especially for the
artificial images with $\beta <1$, where bright sources dominate.

\begin{figure}
\plotone{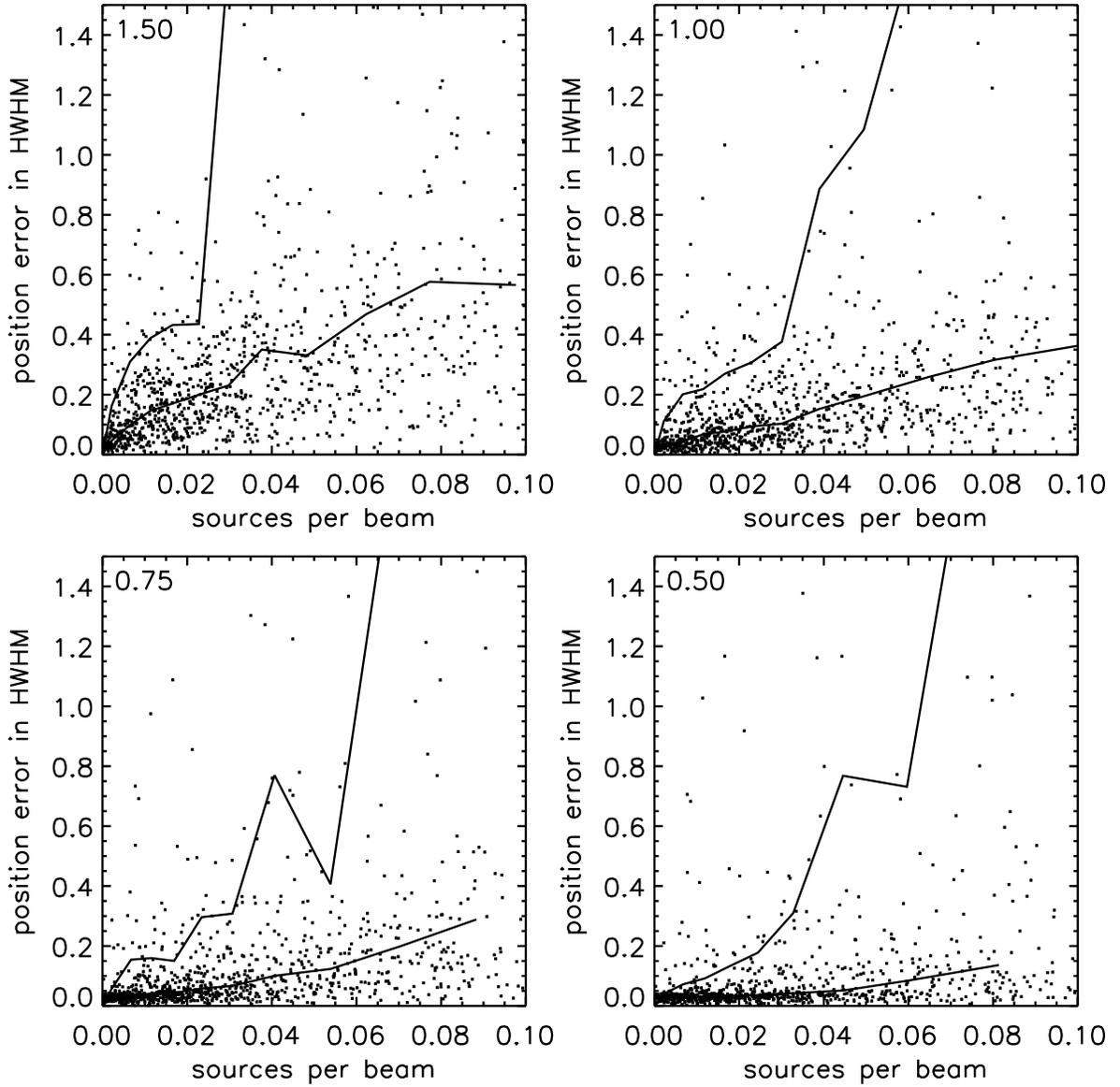}
\caption[ ]{ Astrometry (position) errors as a function of the number
of sources per beam, for isolated sources (see text), given in terms
of the beam HWHM.  These positional errors can be thought-of as really
being as a function of flux, but where for clarity the flux is given
not in Jy but by the source density $s/b$ to that flux level.  The
lower line is a running median and the upper line is a running
90-percent line.  The panels are labeled by number count slope $\beta$
(see text for definition).  The median positional error in the
$\beta=1.5$ case is roughly $0.6\,\theta_{\rm HWHM}$ for sources with
fluxes such that a catalog to that level would contain $s/b=1/30$ of a
source per beam.}
\label{fig:positions}
\end{figure}

The centroid errors for the isolated sources, in units of the beam's
half-width at half maximum (HWHM) are shown in
Figure~\ref{fig:positions} as a function of the source density, or
number of sources per beam $s/b$.  Really these are measured as a
function of flux, but since source density increases with decreasing
flux limit, the quantities are interchangeable; each point is plotted
at the source density $s/b$ which would be found in a source catalog
made down to a limit equal to that source's flux.  A running median
and a running 90~percent level are also shown.  The results are
dramatic.  For $\beta=1.5$, which is the typical count slope for
submillimeter sources or nearby stars, measured source positions are
occasionally (10 percent of the time) displaced from their true source
positions by a significant fraction of the half-power point of the
beam at $s/b=1/40$; such displacements are common at $s/b=1/20$.
These problems are alleviated as the counts become less steep; at
$\beta\sim 0.75$ (the typical count-slope for faint visual and
infrared sources), the worst 10 percent of positions are displaced by
the HWHM at $s/b= 1/17$.  Recall that these numbers (and
Figure~\ref{fig:positions}) are computed only for isolated sources, as
defined above.

Crowding-induced centroid shifts have been observed in some
microlensing events: As a faint lensed star becomes brighter, its
apparent position shifts towards its true position because confusion
noise becomes less important (Goldberg 1998; Goldberg \& Wo\'zniak
1998).

\begin{figure}
\plotone{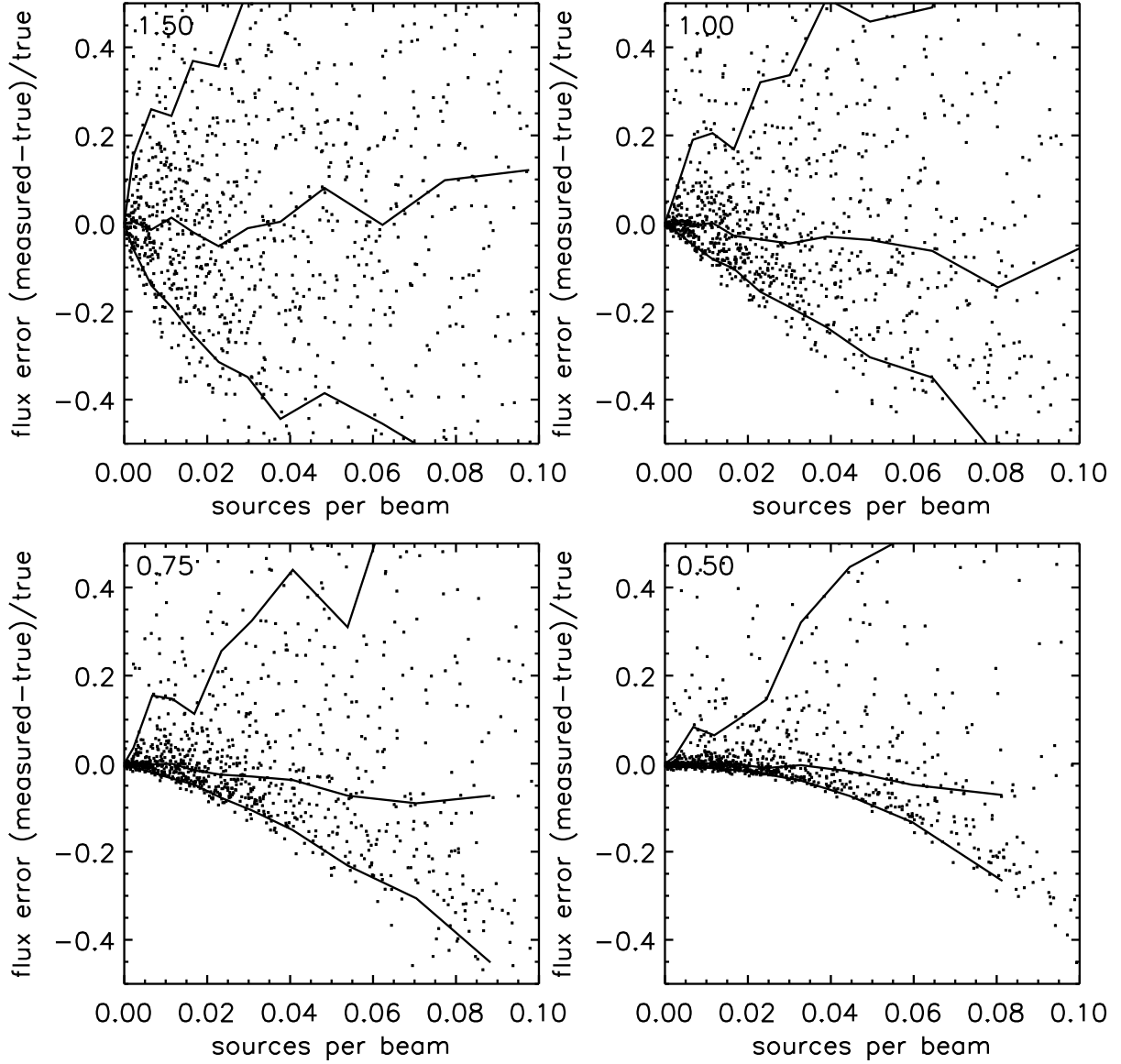}
\caption[ ]{ Fractional flux errors, for isolated sources (see text),
as a function of the number of sources per beam.  The middle line is a
running median and the upper and lower lines are running 10 and
90-percent lines.  The panels are labeled by number count slope
$\beta$ (see text for definition).  The median in the large-$\beta$
figures appears above the majority of the points because there are a
significant number of detected sources with no true source within
1.5\,HWHM.}
\label{fig:fluxes}
\end{figure}

For each measured source, the gaussian beam shape is fit to the peak
in a square box of side length $2\,\theta_{\rm FWHM}$ centered on the
measured centroid.  This provides a measured flux.  The measured and
true fluxes are compared in Figure~\ref{fig:fluxes}.  Again the
results are shown in terms of source density rather than flux level.
The flux errors become bad at roughly the same source per beam levels
as the position errors.  Note also that at the faint end there is a
bias in the median flux error, caused by the subtraction of a finite
background from the images.  This bias will exist in all observations
that are analyzed after background subtraction (all visual and
near-infrared images) and all observations made with chopping (many
infrared and submillimeter images).

\section{Discussion of assumptions}

As has been emphasized, this study makes use of very optimistic
assumptions about the properties of the imaging data.  An experiment
was performed to test one of these assumptions: the perfectly gaussian
beam.  An image was made by ``beam switching'' with a nod of
$3\,\theta_{\rm FWHM}$ so that the beam consists of a central positive
gaussian flanked by two negative gaussians of half the power but the
same FWHM separated by angles of $\pm 3\,\theta_{\rm FWHM}$ on either
side.  These parameters were chosen to roughly match typical
submillimeter observing strategies (eg, Eales et al 1999).  In these
artificial beam-switched images, at constant source density, the
median positional errors are typically $\sim 30$~percent worse at
$\beta= 1.5$ and a factor of $\sim 3$ worse at $\beta=0.75$, relative
to the images made with the single gaussian beam.  Pure gaussian beam
may therefore be an unrealistically optimistic assumption, although it
is close to correct for atmosphere-limited visual and near-infrared
observations.

The assumption of point-like or non-extended sources is overly
optimistic in ground-based and space-based optical imaging, where
recent data is not, by and large, confusion limited.  It is probably
not a problem for recent submillimeter observations, which have a beam
with $\theta_{\rm FWHM}\approx 15$~arcsec.  Unfortunately, a proper
treatment of the effects of finite source sizes involves modeling
distributions of sizes, radial profiles and shapes, all as a function
of flux; this is outside the scope of this work.

The assumption that sources are unclustered is probably optimistic for
virtually all deep imaging observations.  Clustering becomes important
whenever the angular correlation length is larger than or on the order
of the beam size (eg, Barcons et al 1992), which is true for virtually
all optical and near-infrared imaging.  This condition is probably
also met for the submillimeter sources, although at present the
numbers of sources are too small for a direct measurement.  Again, a
full treatment requires parameterization of the angular clustering and
its dependence on flux.

The assumption of power-law number--flux relation must be incorrect in
detail, in the sense that the integrated flux from sources in a power
law diverges either at the faint or bright end (or both).  In
particular, the $\beta=1.5$ models presented here diverge in terms of
total flux (although not in terms of fluctuations) at the faint end.
Experiments of varying the depth to which the artificial source
catalogs go have not shown significant changes in the error
distributions.  The errors are not dominated by the very faintest
sources; they are dominated by the sources with fluxes which fall
between the flux of the source in question and the level at which
there is $s/b\sim 1$ source per beam.  It is the number--flux relation
in this region only which is important to the confusion noise.

\section{Detected sources}

Perhaps the single most unrealistic assumption is that the true source
positions are known in advance; ie, the measurement of centroid
shifts, above, was performed by taking centroids in the vicinities of
the true source positions.  This assumption is extremely optimistic,
because in a real astronomical project, sources are usually detected
ab initio, with no prior knowledge of their positions.

\begin{figure}
\plotone{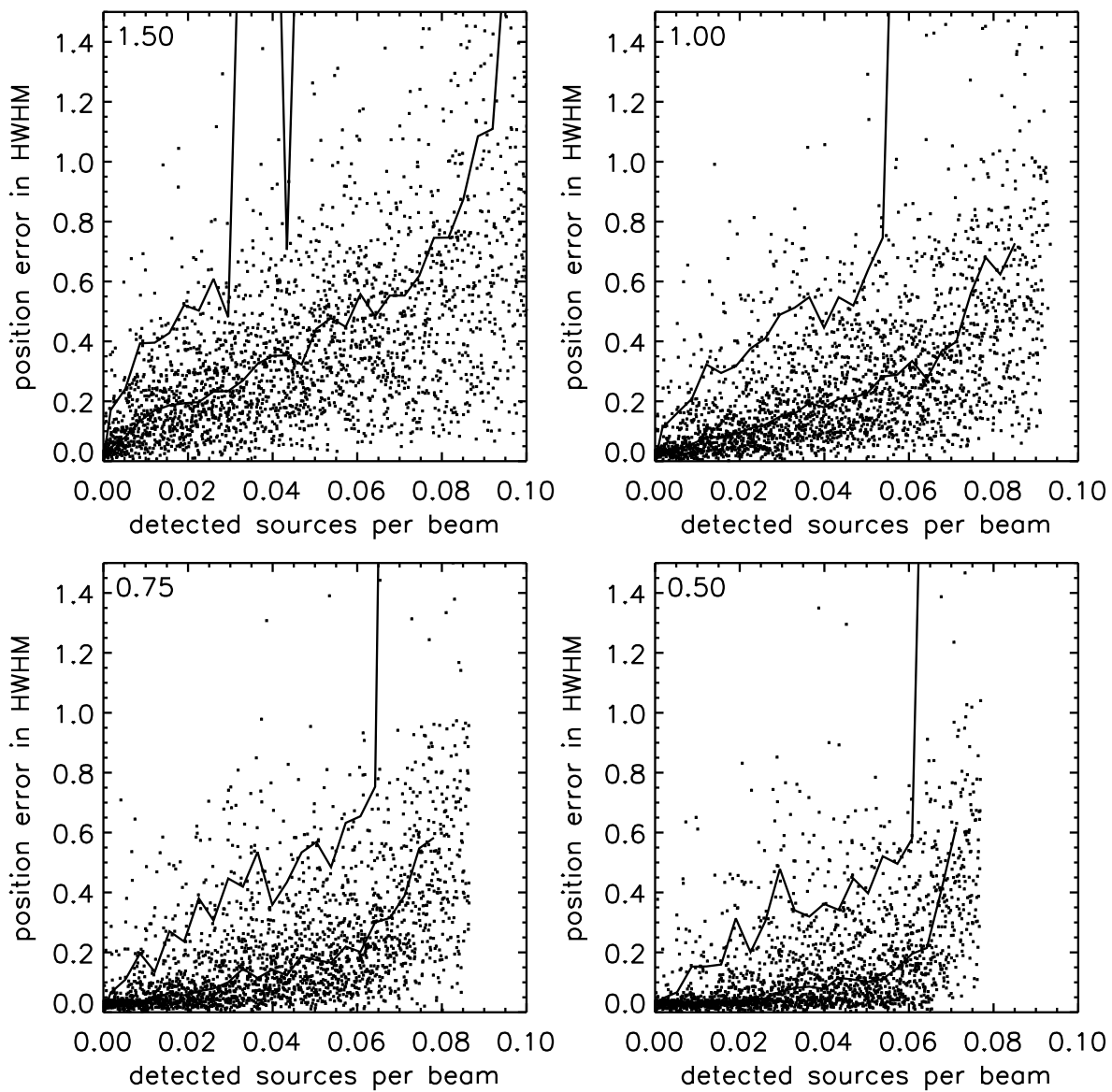}
\caption[ ]{ Astrometry (position) errors as a function of the number
of detected sources per beam, given in terms of the beam HWHM, as in
Figure~\ref{fig:positions}, but now for sources detected with no a
priori knowledge of their positions.  The detected sources were
matched to true source positions using flux cuts; see text for
details.  The lower line is a running median and the upper line is a
running 90-percent line.  The panels are labeled
by true number count slope $\beta$ (see text for definition).}
\label{fig:repositions}
\end{figure}

To test the influence of this optimistic assumption, sources were
detected in the the simulated images with DAOPHOT (Stetson 1987) and
matched, after detection, to the true source positions.  This matching
is not unique; since position shifts are large, there are many faint
``detected'' sources which could potentially be matched with each of
the faint ``true'' source positions, and vice versa.  To reduce this
ambibuity, the detected source flux can be compared with the true
flux.  Figure~\ref{fig:repositions} shows the positional errors
between detected and true sources as a function of detected source
density.

The detected sources were matched to the true source positions by
taking, for each detected source, the closest true source with its
flux between 0.67 and 1.33 of the detected source flux.  This choice
is admittedly arbitrary, but it represents a cut equivalent to
something like 3-sigma.  At bright limits, the flux cut does not
affect the results, but at faint levels, where for every detected
source there are several true source candidates, this cut does affect
the positional errors.  Figure~\ref{fig:repositions} shows that the
positions obtained by detection with no prior information are indeed
worse than those obtained with the a priori knowledge.  This further
strengthens the statement that the confusion effects shown in this
paper are in fact far less severe than in any real astronomical
experiment.

\section{Discussion of recent data}

The deepest recent ground-based visual and near-infrared observations
of blank sky are not confusion limited (eg, Djorgovski et al 1995;
Hogg et al 1997).  However, since they are in the region of $s/b=1/50$
sources per beam, it does not make sense to perform deeper imaging
until wide-field, ground-based adaptive optics can be used.  Radio
imaging has been confusion limited for some time (eg, Condon 1974) so
investigators are usually careful to truncate analyses before
confusion noise becomes destructive.  In all these fields of
astronomy, telescope time is better spent increasing field area than
depth.  Although until the launch of Chandra essentially all deep
x-ray imaging was confusion limited (eg, Barcons 1992), present-day
space-based x-ray and optical imaging is not yet at the confusion
limit (Williams et al 1996; Brandt et al in preparation).  This may
change with planned future instrumentation and exposures.

Unfortunately, several recent publications on faint mid-infrared and
submillimeter sources have ignored confusion as a possible source of
error and are beyond the confusion limit.  This is particularly
serious since the number count slopes are very steep ($2.0<\beta
<1.5$) both observationally and according to simple models (Hughes et
al 1998; Blain et al 1998; Barger et al 1999a; Aussel et al 1999;
Elbaz et al 1999).  One $850~{\rm \mu m}$ study shows sources to
$s/b\sim 1/11$ (Hughes et al 1998) and others show sources to $s/b\sim
1/50$ (Smail et al 1998; Eales et al 1999).  The ISO counts at
$15~{\rm \mu m}$ have been pushed to $s/b\sim 1/25$ (Aussel et al
1999).  The fact that a significant fraction of the submillimeter
sources show no striking visual counterparts is not at all surprising;
the submillimeter positions will be shifted from their true positions
by more than the HWHM or 7.4~arcsec.  The authors generally consider
only a region of radius $\sim \theta_{\rm HWHM}/(S/N)$ where $(S/N)$
is the estimated signal-to-noise ratio of the detection; these radii
are generally 2 to 4 arcsec (Hughes et al 1998; Smail et al 1998;
Barger et al 1999b).  The results of the analysis presented here
suggests that these authors should be looking in a region a factor of
4 to 10 larger in solid angle.

It might be hoped that confusion is not so destructive because perhaps
the true mid-infrared and submillimeter source number--flux relations
are not nearly as steep as what is measured and modeled.  However, if
so, many of the faintest reported sources must truly be spurious.  One
phenomenological model (Barger et al 1999a) shows the number--flux
relation flattening just below the faintest detected sources, but it
does not flatten quickly enough to solve the confusion problem.
Future imaging efforts would be better spent increasing field area
than exposure times, and counterparts ought to be sought in large
error boxes.

A recent comprehensive review (Blain et al 1998) estimates the flux
levels at which future ground- and space-based infrared through radio
surveys will become confusion limited, using the incorrect criterion
of $s/b=1$ source per beam.  Their limiting flux levels become more
realistic when multiplied by a factor of $\sim 10$ to $30$ to bring
the surface densities down to the confusion noise limits presented
here.

Another area in which deep imaging in crowded fields is necessary is
studies of the Galactic center (eg, Eckart \& Genzel 1997; Genzel et
al 1997; Ghez et al 1998).  In these studies, accurate astrometry is
needed not just for identifying counterparts at other wavelengths, but
also for measuring proper motions, on the basis of which the central
black hole mass is estimated.  Current analyses of the Galactic center
go to more crowded depths than $s/b=10$ sources per beam in the
central $1~{\rm arcsec^2}$.  It may be important for the Galactic
center investigators to show that the large proper motions they
observe are truly the motions of individual bright stars and not
seriously affected by many small motions in the underlying sea of
unresolved sources.  If the detected motions have a significant
confusion-induced component, it can be predicted that the stellar
accelerations will deviate significantly from their gravitational
expectations.  This prediction may already have been falsified, at
least at bright levels (Ghez et al 2000).

\section{Conclusions}

For typical faint imaging in the visual and near-infrared, in which
number counts have the form $d\log N/d\log S=S^{-\beta}$ with
$\beta\approx 0.75$, the confusion limit rule-of-thumb that imaging
should not be pursued much fainter than $s/b\sim 1/30$ sources per
beam is essentially correct, both for obtaining good positions and
good photometry.  Optimistic simulations show that positions and
fluxes of sources more numerous than this condition are likely to have
large uncertainties.  When number counts are steep, with Euclidean
$\beta=1.5$ or steeper, the problem is even more severe and a better
rule-of-thumb is something like $s/b\sim 1/50$.

Source identifications in one set of imaging data based on detections
in another set will be affected by these confusion-induced astrometry
errors.  It is essential that surveys working near the confusion limit
perform realistic simulations (which include the sources well faint of
any detection limits) in order to draw conservative positional error
boxes for source identification.

\acknowledgements Gerry Neugebauer and Tom Soifer drummed the ``30
beams per source'' rule-of-thumb into my head.  Useful comments, code
and information came from Tal Alexander, John Bahcall, Roger
Blandford, Tom Chester, Judy Cohen, Jim Condon, Daniel Eisenstein, Tom
Jarrett, Wayne Landsman, Bruce Partridge, Eric Richards, Douglas Scott
and Ian Smail.  Financial support was provided under Hubble Fellowship
grant HF-01093.01-97A from STScI, which is operated by AURA under NASA
contract NAS~5-26555.  This research made use of the NASA ADS Abstract
Service.

\end{document}